\documentclass[aps,prl,twocolumn,amsmath,amssymb,showpacs]{revtex4}

\def\Tr{\mbox{Tr}}

\begin{document}

\title{Order Symmetry of Weak Measurements}

\author{Lars M. Johansen}
\affiliation{Department of Technology, Buskerud University College,
N-3601 Kongsberg, Norway}
\email{lars.m.johansen@hibu.no}

\author{Pier A. Mello}
\affiliation{
Instituto de F\'{\i}sica, Universidad Nacional Aut\'{o}noma de M\'{e}xico,
Ap. Postal 20-364, 01000 M\'{e}xico, D.F., Mexico}
\email{mello@fisica.unam.mx}

\begin{abstract}

Weak values are usually associated with weak measurements of an observable on a pre- and post-selected ensemble. We show that more generally, weak values are proportional to the correlation between two pointers in a successive measurement. We show that this generalized concept of weak measurements displays a symmetry under reversal of measurement order. We show that the conditions for order symmetry are the same as in classical mechanics. We also find that the imaginary part of the weak value has a counterpart in classical mechanics. This scheme suggests new experimental possibilities.

\end{abstract}



\maketitle

In recent years, there has been a surge of interest in generalized forms of quantum measurements. This has been driven by recent technological progress in fields such as quantum optics and quantum information theory. The standard type of measurement, a projective measurement, requires a strong interaction between the measurement apparatus (or pointer) and the system \cite{pointer}. In particular, the measurement interaction should influence the pointer so strongly that different eigenstates of the measured observable lead to orthogonal pointer states. On the other hand, in a weak measurement, the interaction between pointer and system hardly affects the pointer.

Of particular interest is the subject of weak measurements of an observable, say $\hat{A}$, on a system pre-selected (or prepared) in a specific state, say $| \psi \rangle$, and post-selected in another state, say $| \phi \rangle$. The average reading of the position of the pointer that measures $\hat{A}$ in such an experiment is the real part of the quantity \cite{Aharonov+AlbertETAL-ResuMeasCompSpin:88}
\begin{equation}
    A_w = {\langle \phi | \hat{A} | \psi \rangle \over \langle \phi | \psi \rangle} = {\langle \psi | \mathbb{P} \hat{A} | \psi \rangle \over \langle \psi | \mathbb{P} | \psi \rangle},
    \label{eq:weakvalue}
\end{equation}
where $\mathbb{P} = | \phi \rangle \langle \phi |$. The quantity $A_w$, usually complex, is called the ``weak value'' of $\hat{A}$ for the given pre- and post-selected state. The imaginary part of this quantity may also be observed in a related experiment \cite{Aharonov+AlbertETAL-ResuMeasCompSpin:88}. The weak value of an observable may differ considerably from the eigenvalues. In particular, the real part may lie well outside the eigenvalue spectrum \cite{Aharonov+AlbertETAL-ResuMeasCompSpin:88}.
These phenomena are of a purely quantum mechanical origin and would not be expected in a model obeying classical statistics. Such ``strange'' weak values can be observed in a variety of systems \cite{Aharonov+Vaidman-CharQuanSystBetw:89,Aharonov+PopescuETAL-MeasErroNegaKine:93}. Weak measurements of weak values throw light on some of the most non-intuitive aspects of quantum mechanics such as Hardy's paradox \cite{Aharonov+BoteroETAL-ReviHardpara:02,Lundeen+Steinberg-ExpeJoinMeasPhot:09}. They offer a new way of amplifying weak effects \cite{Aharonov+AlbertETAL-ResuMeasCompSpin:88,Hosten+Kwiat-ObseSpinHallEffe:08}.
Weak values may also be observed in projective measurements \cite{Johansen-Recoweakvaluwith:07}.
Furthermore, weak values may provide a complete representation of quantum states \cite{Johansen-Quantheosuccproj:07}.
It has been found that weak values have applications in areas as diverse as e.g. cavity-QED experiments  \cite{Wiseman-Weakvaluquantraj:02}, optical telecom networks \cite{Brunner+AcinETAL-OptiTeleNetwWeak:03}, measuring group velocities \cite{Solli+McCormickETAL-FastLighSlowLigh:04,Brunner+ScaraniETAL-DireMeasSupeGrou:04},
the Leggett-Garg inequality \cite{Williams+Jordan-WeakValuLeggIneq:08} and amplification of weak signals \cite{Ritchie+StoryETAL-RealMeasWeakValu:91,Hosten+Kwiat-ObseSpinHallEffe:08}. The theory has been confirmed in a number of experiments \cite{Ritchie+StoryETAL-RealMeasWeakValu:91,Parks+CullinETAL-ObsemeasoptiAhar:98, Brunner+AcinETAL-OptiTeleNetwWeak:03,Solli+McCormickETAL-FastLighSlowLigh:04, Pryde+OBrienETAL-MeasQuanWeakValu:05,Wang+SunETAL-Expedemomethreal:06,
Hosten+Kwiat-ObseSpinHallEffe:08,Lundeen+Steinberg-ExpeJoinMeasPhot:09}.

Weak values result from a weak measurement of an observable followed by a post-selection. In a certain sense, the weak value is a conditional expectation of the observable. However, there is a specific measurement order involved: the observable is measured before the condition. This begs the question: are weak values unavoidably connected to a specific measurement order? Put differently: are weak values unavoidably tied up with pre- and post-selected ensembles?

We thus come to the main purpose of this letter. We first show that weak values may be associated with the correlation between {\em two pointers} in a successive measurement. One pointer measures an observable, while the other measures a projector. We then show that this generalized concept of weak measurements displays a symmetry under reversal of the measurement order: reversing the measurement order gives rise to the complex conjugate of the weak value. We therefore shall see that weak values may very well be observed in weak measurements of reverse order, i.e., measurements where the condition is measured prior to the observable. But again, the requirement is that the first measurement be sufficiently weak. In such experiments there is no selection of a subensemble, because the initial projector measurement is so weak that the eigenvalues of the projector cannot be distinguished. On the other hand, the eigenvalues of the observable itself may readily be observed. It seems to us that this scheme opens up a wide range of new experimental possibilities.

In classical mechanics, measurement outcomes are usually
taken to be independent of the order in which the
measurements are performed. As will be shown here,
this need not always be the case. There are certain conditions
that must be imposed on the measurement pointer
in order for a weak measurement in classical mechanics
to be order independent. We find that these conditions
are the same that must be imposed on the pointer in a
weak quantum measurement for order independence to
be established. Furthermore, we find that weak values
also may be defined in classical mechanics. Surprisingly,
we find an equivalent of the imaginary part of the weak
value in classical mechanics. In fact, the classical theory
of weak values emerge by reducing commutators to Poisson
brackets and anticommutators to twice the product
of observables. This dequantization needs to be done both on
the system proper and on the first pointer.

In the standard treatment of weak values \cite{Aharonov+AlbertETAL-ResuMeasCompSpin:88}, von Neumann's measurement model is used to represent the interaction between the system and a pointer
(the latter being defined in terms of continuous dynamical variables).
It should be noted that weak values are not explicitly tied to the von Neumann model of measurement. Weak values also follow from other types of measurement models. For example, weak values may be observed using a pointer observable with
a finite-dimensional discrete spectrum (see, e.g. \cite{Brun+DiosiETAL-Testweakmeastwo-:08}). Weak values also may be derived using the theory of effects and operations \cite{Garretson+WisemanETAL-uncerelawhicexpe:04}.

Historically, only {\em one} pointer is introduced for the observable $\hat{A}$ being measured first, the post-selection being trivially described by orthodox quantum measurement theory \cite{Aharonov+AlbertETAL-ResuMeasCompSpin:88}.
Here, we shall treat the two observables $\hat{A}$ and $\hat{B}$ symmetrically. Therefore, we introduce {\em two} pointers, $M_1$ and $M_2$, for measuring
in succession the observables $\hat{A}$ and $\hat{B}$, respectively.
In a recent publication \cite{Johansen+Mello-QuanMechSuccMeas:08}, a formalism was developed for the study of successive measurements with an arbitrary interaction strength with the pointers.
Consider the successive measurement of two arbitrary observables, $\hat{A} = \sum_n a_n \mathbb{P}_{a_n}$ and $\hat{B} = \sum_m b_m \mathbb{P}_{b_m}$, where
$\mathbb{P}_{a_n}$ and $\mathbb{P}_{b_m}$ are eigenprojectors corresponding to the eigenvalues $a_n$ and $b_m$, respectively. We assume that $\hat{A}$ is measured before $\hat{B}$. The eigenprojectors satisfy the orthogonality relations
$\mathbb{P}_{a_n} \mathbb{P}_{a_{n'}}  = \delta_{n n'}  \mathbb{P}_{a_n}$,
$\mathbb{P}_{b_m} \mathbb{P}_{b_{m'}}  = \delta_{m m'}  \mathbb{P}_{b_m}$,
as well as the completeness relations $\sum_n \mathbb{P}_{a_n} = \hat{1}$,
$\sum_m \mathbb{P}_{b_m}  = \hat{1}$.
For the analysis of successive measurements, the standard von Neumann measurement model
\cite{Neumann-MathFounQuanMech:55,Aharonov+AlbertETAL-ResuMeasCompSpin:88}
is generalized to let the pointer $M_1$ interact before the pointer $M_2$,
thus giving the interaction Hamiltonian
\begin{equation}
\hat{V} (t) = \epsilon_1 \delta (t-t_1) \hat{A} \hat{P}_1
+ \epsilon_2 \delta (t-t_2) \hat{B} \hat{P}_2, \;\;\;\;\;
0 < t_1 < t_2 .
\label{eq:hamiltonian}
\end{equation}
Here, $\epsilon_i$ $(i=1,2)$ is the strength of the interaction between the system and pointer $M_i$, $t_i$ is the time at which there is an impulsive interaction between the pointer $M_i$ and the system, and $\hat{P}_i$ is the momentum observable for pointer $M_i$.

We assume that prior to the measurement interaction the state of the system is $\hat{\rho}_s$, the state of pointer $M_i$  is $\hat{\rho}_{M_i}$, and the state of the complete experimental arrangement is $\hat{\rho} =  \hat{\rho}_{s} \hat{\rho}_{M_1} \hat{\rho}_{M_2}$. One then finds that the full state after $t_2$ is \cite{Johansen+Mello-QuanMechSuccMeas:08}
\begin{eqnarray}
&& \hat{\rho}^{(\hat{B} \leftarrow \hat{A})} = \sum_{nn'mm'}
(\mathbb{P}_{b_{m}} \mathbb{P}_{a_{n}}  \hat{\rho}_{s} \; \mathbb{P}_{a_{n'}} \mathbb{P}_{b_{m'}})
\nonumber \\
&\times&  \left(
e^{-i\epsilon_1 a_n \hat{P_1}} \hat{\rho}_{M_1}
e^{i\epsilon_1 a_{n'}\hat{P_1}} \right)
\left(
e^{-i\epsilon_2 b_m \hat{P_2}}  \hat{\rho}_{M_2}
e^{i\epsilon_2 b_{m'} \hat{P_2}}
\right ),
\hspace{5mm}
\label{rho t1<t<t2 12 txt}
\end{eqnarray}
where the superscript $(\hat{B} \leftarrow \hat{A})$ indicates that that this is the state after both pointers have interacted with the system.

For a sufficiently strong interaction $\epsilon_1$, the position $\hat{Q}_1$ of pointer 1 will register the eigenvalues of the observable $\hat{A}$ \cite{Neumann-MathFounQuanMech:55}. We will be interested in the opposite case, i.e., a weak interaction where $\epsilon_1$ is small.
Also, for a sufficiently strong interaction $\epsilon_2$, the pointer $\hat{Q}_2$ will register the eigenvalues of the observable $\hat{B}$. However, the distribution of the eigenvalues of $\hat{B}$ may be disturbed by the preceding measurement of $\hat{A}$. In the standard weak measurement scheme \cite{Aharonov+AlbertETAL-ResuMeasCompSpin:88}, also the momentum $\hat{P}_1$ of the first pointer is involved to register the imaginary part of the weak value. For these reasons, we shall first consider the  general case where an arbitrary observable $\hat{F}_1 \equiv F(Q_1,P_1)$ is observed on pointer $M_1$. Furthermore, we assume that $\hat{Q}_2$ is observed on pointer $M_2$.
Next, we calculate the correlation function
\begin{eqnarray}
 \left\langle \left[ \hat{F}_1 - Tr_{M_1}(\rho_{M_1}\hat{F}_1)\right] \, \hat{Q}_2  \right\rangle
    ^{(\hat{B} \leftarrow \hat{A})}
= \sum_{nn'm} \Tr_s (\hat{\rho}_{s} \; \mathbb{P}_{a_{n'}} \mathbb{P}_{b_{m}} \mathbb{P}_{a_{n}})
\nonumber \\
\times  \Tr_1
\left\{
e^{-i\epsilon_1 a_n \hat{P_1}} \hat{\rho}_{M_1}
e^{i\epsilon_1 a_{n'}\hat{P_1}}
\left[ \hat{F}_1 - Tr_{M_1}(\rho_{M_1}\hat{F}_1)\right]
\right\}
\nonumber \\
\times \Tr_2 \left(
e^{-i\epsilon_2 b_m \hat{P_2}}  \hat{\rho}_{M_2}
e^{i\epsilon_2 b_{m} \hat{P_2}} \hat{Q}_2 \right ).
\hspace{5mm}
\end{eqnarray}
Now we assume that the first measurement is weak, so that we expand to first order in $\epsilon_1$.
Assuming also that
\begin{subequations}
\begin{eqnarray}
\Tr_{M_2}(\rho_{M_2} \hat{Q}_2) &=& 0,
\label{<Q2>2=0}
\\
\Tr_{M_1}(\rho_{M_1} \hat{P}_1) &=& 0 \; ,
\label{<P1>1=0}
\end{eqnarray}
\label{<Q2>2=<P1>1=0}
\end{subequations}

we obtain
\begin{eqnarray}
\left\langle
\left[ \hat{F}_1 - Tr_{M_1}(\rho_{M_1}\hat{F}_1)\right]\, \hat{Q}_2  \right\rangle^{(\hat{B} \leftarrow \hat{A})}
\nonumber\\
= i \epsilon_1 \epsilon_2 \left [ \Tr_s (\hat{\rho}_s \hat{A} \hat{B}) \Tr_1 ( \hat{\rho}_{M_1} \hat{P}_1 \hat{F}_1) \right. \nonumber \\ - \left. \Tr_s (\hat{\rho}_s \hat{B} \hat{A}) \Tr_1 ( \hat{\rho}_{M_1} \hat{F}_1 \hat{P}_1) \right ] + \mathcal{O}(\epsilon_1^2).
\end{eqnarray}
Thus, we have
\begin{eqnarray}
\lim_{\epsilon_1 \rightarrow 0}
\frac{1}{\epsilon_1 \epsilon_2}
\left\langle \left[ \hat{F}_1 - Tr_{M_1}(\rho_{M_1}\hat{F}_1)\right] \, \hat{Q}_2  \right\rangle^{(\hat{B} \leftarrow \hat{A})}
\nonumber \\
= {i \over 2} \left [ \Tr_s \left ( \hat{\rho}_s \{ \hat{A},\hat{B} \} \right ) \Tr_1 \left ( \hat{\rho}_{M_1} [\hat{P}_1, \hat{F}_1 ] \right ) \right. \nonumber \\ + \left. \Tr_s \left ( \hat{\rho}_s [\hat{A},\hat{B}] \right ) \Tr_1 \left ( \hat{\rho}_{M_1} \{ \hat{P}_1, \hat{F}_1 \} \right ) \right ] \; ,
\label{eq:generalcorrelation}
\end{eqnarray}
where $[ \hat{A},\hat{B}]=\hat{A} \hat{B} - \hat{B}\hat{A}$
and $\{ \hat{A},\hat{B} \}= \hat{A} \hat{B} + \hat{B}\hat{A}$
denote the commutator and anticommutator of $\hat{A}$ and $\hat{B}$, respectively. A sufficient condition for symmetry under order exchange $\hat{A} \leftrightarrow \hat{B}$ is
\begin{equation}
    \Tr_1 \left ( \hat{\rho}_{M_1} \{ \hat{P}_1, \hat{F}_1 \} \right ) =0,
    \label{eq:symmetry}
\end{equation}
whereas a sufficient condition for anti-symmetry is
\begin{equation}
    \Tr_1 \left ( \hat{\rho}_{M_1} [\hat{P}_1, \hat{F}_1 ] \right ) = 0.
    \label{eq:antisymmetry}
\end{equation}
One possible way of fulfilling (\ref{eq:antisymmetry}) is $[\hat{P}_1, \hat{F}_1 ]=0$. This is ensured provided that $\hat{F}_1 = \hat{F}_1(\hat{P}_1)$. A sufficient condition for symmetry (\ref{eq:symmetry}) is that $F_1=F_1(\hat{Q}_1)$ and that the current density of the first pointer vanishes \cite{Johansen-WeakMeaswithArbi:04},
\begin{equation}
    \langle Q_1 | \hat{\rho}_{M_1} \hat{P}_1 + \hat{P}_1 \hat{\rho}_{M_1} | Q_1 \rangle =  0.
    \label{eq:vanishingcurrent}
\end{equation}
For commuting observables, $[\hat{A}, \hat{B}] = 0$, symmetry is assured even for nonvanishing current density.

Thus, we see that the average of the symmetric and antisymmetric product of two observables can be obtained as the correlation between two pointers in a successive measurement, provided that the first measurement is weak \cite{Johansen+Mello-QuanMechSuccMeas:08}. Note that there is no condition on the interaction strength $\epsilon_2$ of the second pointer.

Furthermore, it is easily shown that
\begin{subequations}
\begin{eqnarray}
\frac{\langle Q_1 \rangle}
{\epsilon_1}^{(\hat{B} \leftarrow \hat{A})}  &=&   \Tr_s (\hat{\rho}_{s} \hat{A} ),
    \label{Q1single}
\\
        \lim_{\epsilon_1 \rightarrow 0}
\frac{\langle \hat {Q}_2 \rangle}
{\epsilon_2}^{(\hat{B} \leftarrow \hat{A})}
&=&   \Tr_s (\hat{\rho}_{s} \hat{B} ),
    \label{Q2single}
\end{eqnarray}
\label{single}
\end{subequations}
We have assumed
\begin{equation}
\Tr(\rho_{M_1}Q_1) =0 .
\label{<Q1>1=0}
\end{equation}
in deriving Eq. (\ref{Q1single}) and Eq. (\ref{<P1>1=0}) in deriving Eq. (\ref{Q2single}). Eq. (\ref{Q1single}) is valid regardless of the interaction strength $\epsilon_1$, while Eq. (\ref{Q2single}) holds only for weak interaction.

Now we examine the possibility of observing weak values as a correlation between the two pointers. We are motivated by Eq. (\ref{eq:weakvalue}), which tells us that at least in the case of commuting observables
$\hat{A}$ and $\mathbb{P}$, a weak value may be understood as a conditional correlation between the observable and the projector.
First, we consider the case where the second observable is a projector,
\begin{equation}
    \hat{B} =  \mathbb{P},
\end{equation}
where $\mathbb{P}^2=\mathbb{P}$.
Under the conditions (\ref{eq:vanishingcurrent}), (\ref{<Q1>1=0}) and (\ref{<P1>1=0}),
we obtain from Eqs. (\ref{eq:generalcorrelation}) and (\ref{Q2single}) that
\begin{subequations}
\begin{eqnarray}
 \lim_{\epsilon_1 \rightarrow 0}  \left [
\frac{\langle Q_1  Q_2 \rangle^{(\mathbb{P} \leftarrow \hat{A})}}{\epsilon_1\langle Q_2 \rangle^{(\mathbb{P} \leftarrow \hat{A})}}
\right ]
&=& \Re  \left [ \frac{\Tr_s (\hat{\rho}_{s}
\mathbb{P} \hat{A} )}{\Tr_s (\hat{\rho}_{s} \mathbb{P})} \right ],
\label{eq:projectorfirstreal}\\
  \lim_{\epsilon_1 \rightarrow 0}  \left [
\frac{\langle  P_1  Q_2 \rangle^{(\mathbb{P} \leftarrow \hat{A})}}{\epsilon_1\langle Q_2 \rangle^{(\mathbb{P} \leftarrow \hat{A})}}
\right ] &=&  2 \sigma_{P_1}^2
\Im  \left [ \frac{\Tr_s (\hat{\rho}_{s}
\mathbb{P} \hat{A})} {\Tr_s (\hat{\rho}_{s} \mathbb{P})} \right ], \qquad
        \label{eq:projectorfirstimaginary}
\end{eqnarray}
        \label{eq:projectorfirst}
\end{subequations}
where
\begin{equation}
    \sigma_{P_1}^2 = \Tr \hat{\rho}_{M_1} (\hat{P}_1)^2.
\end{equation}
We see from Eq. (\ref{eq:weakvalue}) that these conditional correlations measure the real and imaginary parts of the weak value of $\hat{A}$.
Note that the result is independent of the strength $\epsilon_2$ of the second measurement.

But we may as well consider the reverse order measurement, i.e., one where the projector is measured first,
\begin{equation}
    \hat{A}= \mathbb{P}.
\end{equation}
In considering the reverse order measurement, we still assume that the system first interacts with pointer $M_1$ and then with pointer $M_2$.
In this case, it follows from Eq. (\ref{eq:generalcorrelation}) under the conditions (\ref{eq:vanishingcurrent}), (\ref{<Q1>1=0}) and (\ref{<P1>1=0}) that
\begin{subequations}
\begin{eqnarray}
        \lim_{\epsilon_1 \rightarrow 0}
\left [
\frac{\langle Q_1 Q_2 \rangle^{(\hat{B} \leftarrow \mathbb{P} )}}
{\epsilon_2 \langle Q_1 \rangle^{(\hat{B} \leftarrow \mathbb{P})}}
\right ]
&=& \Re  \left [ \frac{\Tr_s (\hat{\rho}_{s} \hat{B}\mathbb{P} )}{\Tr_s (\hat{\rho}_{s} \mathbb{P})} \right ],
\label{eq:projectorlastreal}
\\
        \lim_{\epsilon_1 \rightarrow 0}  \left [
        \frac{\langle P_1 Q_2 \rangle^{(\hat{B} \leftarrow \mathbb{P} )}}{\epsilon_2 \langle Q_1 \rangle^{(\hat{B} \leftarrow \mathbb{P} )}}
        \right ] &=& 2 \sigma_{P_1}^2
        \Im  \left [ \frac{\Tr_s (\hat{\rho}_{s}  \hat{B} \mathbb{P} )} {\Tr_s (\hat{\rho}_{s} \mathbb{P})} \right ]. \qquad
        \label{eq:projectorlastimaginary}
\end{eqnarray}
        \label{eq:projectorlast}
\end{subequations}
As seen from Eq. (\ref{eq:weakvalue}), in this case we obtain the real and the (negative) imaginary part of the  weak value of $\hat{B}$.
More explicitly, if we perform the replacement $\hat{B} \rightarrow \hat{A}$ in Eqs. (\ref{eq:projectorlast}), comparison with Eqs. (\ref{eq:projectorfirst})
shows that reversal of measurement order gives the complex conjugate of the weak value.
One can show that there is no such symmetry for arbitrary coupling strengths.
Since the interaction strength $\epsilon_1$ with the pointer $M_1$ goes to zero, we cannot identify the eigenvalues of the projector $\mathbb{P}$. Therefore, we have no possibility of separating into two subensembles as in a standard weak measurement. On the other hand, we may choose whatever interaction strength $\epsilon_2$ that we like. For example, for $\epsilon_2 \rightarrow \infty$ we have a projective measurement of $\hat{B}$ \cite{Johansen+Mello-QuanMechSuccMeas:08}. Thus, this experiment permits us to observe both the eigenvalues and the weak value of the observable $\hat{B}$ simultaneously.

It is interesting at this point to compare the above results with a purely classical model \cite{Johansen-WeakMeaswithArbi:04}. The starting point is the interaction
\begin{equation}
V(t) = \epsilon_1 \delta (t-t_1) A(q,p) P_1
+ \epsilon_2 \delta (t-t_2) B(q,p) P_2, \;\;\;\;\;
0 < t_1 < t_2 ,
\end{equation}
where $q$, $p$ represent the phase space variables of the system proper and $Q_i$, $P_i$ represent those of the pointer $M_i$ ($i=1,2$). The initial state is the phase space density
\begin{equation}
    \rho(q,p,Q_1,P_1,Q_2,P_2) = \rho_s(q,p) \rho_{M_1} (Q_1,P_1) \rho_{M_2} (Q_2,P_2).
\end{equation}
and the aim is to find its evolution with time for small $\epsilon_1$. We proceed in a manner analogous to the quantum case. The detailed calculation will be published elsewhere and gives the following result:
\begin{eqnarray}
&&\lim_{\epsilon_1 \rightarrow 0}
\frac{1}{\epsilon_1 \epsilon_2}
\Big\langle
\big[ F_1(Q_1,P_1) - \big\langle F_1(Q_1,P_1)\big\rangle \big] Q_2
\Big\rangle^{(\hat{B} \leftarrow \hat{A})}
\nonumber \\
&&\hspace{5mm}= -\Big[
\langle A B \rangle_s
\big\langle [P_1, F_1(Q_1,P_1) ]_{PB}
\big\rangle_{M_1}
\nonumber \\
&&\hspace{15mm}+ \langle [ A,B ]_{PB} \rangle
\big\langle P_1 F_1(Q_1,P_1) \big\rangle_{M_1}
\Big] \; ,
\label{eq:classical}
\end{eqnarray}
where $[A,B]_{PB}$ indicates a Poisson bracket. Notice that Eq. (\ref{eq:classical}) obtains from Eq. (\ref{eq:generalcorrelation}) by changing commutators to $i\hbar \times$ classical Poisson brackets and anti-commutators to twice the product of the two observables. It is remarkable that under this exchange, the same conditions indicated right after Eq. (\ref{eq:generalcorrelation}) for the symmetry and anti-symmetry of the result with respect to the interchange of the measurement of $A$ and $B$ applies to the present classical case as well.

In conclusion, we have demonstrated how weak values may be associated with the conditional correlation between two pointers in a successive measurement.
The standard scheme for measuring weak values is a special case of this, where the last measurement is a projective measurement of a projector. With the generalized measurement scheme, weak values may be observed in weak measurements with a reverse measurement order. We have considered reverse order measurements in the sense that the post-selection is replaced by an initial weak projector measurement.
As in standard weak measurements, the essential requirement is that the first measurement be weak.
In a reverse order weak measurement, the complex conjugate of the weak value is observed. This is reminiscent of a different form of reverse order weak measurement, where pre-selection and post-selection are interchanged.
Interestingly, as follows directly from Eq. (\ref{eq:weakvalue}), this also gives the complex conjugate weak value.

The possibility of performing weak measurements in reverse order widens considerably the available experimental conditions under which weak values may be observed.

In a weak measurement, we lose access to eigenvalues of the observable measured first due to the weakened interaction. So, in a weak measurement of standard order we do not see the eigenvalues of the observable, but may perform a perfect post-selection. In a weak measurement of reverse order, we have no possibility of distinguishing between the two alternatives of the projector, but here we can obtain the eigenvalues of the observable itself.

We have found that both in a classical as well as in a quantum mechanical description, the same conditions apply for symmetry and antisymmetry of pointer correlations under the exchange of measurement operations. Remarkably, there is a classical counterpart to the imaginary part of the weak value. The physical significance of this fact needs further investigation.

In order to emphasize the symmetry of the situation, the calculations in this paper were made with two pointers. However, it may be noted that an equivalent analysis can be made by representing only the first measurement with a pointer.


\begin{thebibliography}{22}
\expandafter\ifx\csname natexlab\endcsname\relax\def\natexlab#1{#1}\fi
\expandafter\ifx\csname bibnamefont\endcsname\relax
  \def\bibnamefont#1{#1}\fi
\expandafter\ifx\csname bibfnamefont\endcsname\relax
  \def\bibfnamefont#1{#1}\fi
\expandafter\ifx\csname citenamefont\endcsname\relax
  \def\citenamefont#1{#1}\fi
\expandafter\ifx\csname url\endcsname\relax
  \def\url#1{\texttt{#1}}\fi
\expandafter\ifx\csname urlprefix\endcsname\relax\def\urlprefix{URL }\fi
\providecommand{\bibinfo}[2]{#2}
\providecommand{\eprint}[2][]{\url{#2}}

\bibitem{pointer}

In the present paper, we shall use the term ``pointer'' to represent the measurement apparatus. Other equivalent terms often used in the literature are ``probe'' and ``meter''.

\bibitem[{\citenamefont{Aharonov et~al.}(1988)\citenamefont{Aharonov, Albert,
  and Vaidman}}]{Aharonov+AlbertETAL-ResuMeasCompSpin:88}
\bibinfo{author}{\bibfnamefont{Y.}~\bibnamefont{Aharonov}},
  \bibinfo{author}{\bibfnamefont{D.~Z.} \bibnamefont{Albert}},
  \bibnamefont{and} \bibinfo{author}{\bibfnamefont{L.}~\bibnamefont{Vaidman}},
  \bibinfo{journal}{Phys. Rev. Lett.} \textbf{\bibinfo{volume}{60}},
  \bibinfo{pages}{1351} (\bibinfo{year}{1988}).

\bibitem[{\citenamefont{Aharonov and
  Vaidman}(1989)}]{Aharonov+Vaidman-CharQuanSystBetw:89}
\bibinfo{author}{\bibfnamefont{Y.}~\bibnamefont{Aharonov}} \bibnamefont{and}
  \bibinfo{author}{\bibfnamefont{L.}~\bibnamefont{Vaidman}}, in
  \emph{\bibinfo{booktitle}{Bell's Theorem, Quantum Theory and Conceptions of
  the Universe}}, edited by
  \bibinfo{editor}{\bibfnamefont{M.}~\bibnamefont{Kafatos}}
  (\bibinfo{publisher}{Kluwer Academic Publishers}, \bibinfo{year}{1989}), pp.
  \bibinfo{pages}{17--22}.

\bibitem[{\citenamefont{Aharonov et~al.}(1993)\citenamefont{Aharonov, Popescu,
  Rohrlich, and Vaidman}}]{Aharonov+PopescuETAL-MeasErroNegaKine:93}
\bibinfo{author}{\bibfnamefont{Y.}~\bibnamefont{Aharonov}},
  \bibinfo{author}{\bibfnamefont{S.}~\bibnamefont{Popescu}},
  \bibinfo{author}{\bibfnamefont{D.}~\bibnamefont{Rohrlich}}, \bibnamefont{and}
  \bibinfo{author}{\bibfnamefont{L.}~\bibnamefont{Vaidman}},
  \bibinfo{journal}{Phys. Rev. A} \textbf{\bibinfo{volume}{48}},
  \bibinfo{pages}{4084} (\bibinfo{year}{1993}).

\bibitem[{\citenamefont{Aharonov et~al.}(2002)\citenamefont{Aharonov, Botero,
  Popescu, Reznik, and Tollaksen}}]{Aharonov+BoteroETAL-ReviHardpara:02}
\bibinfo{author}{\bibfnamefont{Y.}~\bibnamefont{Aharonov}},
  \bibinfo{author}{\bibfnamefont{A.}~\bibnamefont{Botero}},
  \bibinfo{author}{\bibfnamefont{S.}~\bibnamefont{Popescu}},
  \bibinfo{author}{\bibfnamefont{B.}~\bibnamefont{Reznik}}, \bibnamefont{and}
  \bibinfo{author}{\bibfnamefont{J.}~\bibnamefont{Tollaksen}},
  \bibinfo{journal}{Phys. Lett. A} \textbf{\bibinfo{volume}{301}},
  \bibinfo{pages}{130} (\bibinfo{year}{2002}).

\bibitem[{\citenamefont{Lundeen and
  Steinberg}(2009)}]{Lundeen+Steinberg-ExpeJoinMeasPhot:09}
\bibinfo{author}{\bibfnamefont{J.~S.} \bibnamefont{Lundeen}} \bibnamefont{and}
  \bibinfo{author}{\bibfnamefont{A.~M.} \bibnamefont{Steinberg}},
  \bibinfo{journal}{Phys. Rev. Lett.} \textbf{\bibinfo{volume}{102}},
  \bibinfo{pages}{020404} (\bibinfo{year}{2009}).

\bibitem[{\citenamefont{Hosten and
  Kwiat}(2008)}]{Hosten+Kwiat-ObseSpinHallEffe:08}
\bibinfo{author}{\bibfnamefont{O.}~\bibnamefont{Hosten}} \bibnamefont{and}
  \bibinfo{author}{\bibfnamefont{P.}~\bibnamefont{Kwiat}},
  \bibinfo{journal}{Science} \textbf{\bibinfo{volume}{319}},
  \bibinfo{pages}{788} (\bibinfo{year}{2008}).

\bibitem[{\citenamefont{Johansen}(2007{\natexlab{a}})}]{Johansen-Recoweakvaluw%
ith:07}
\bibinfo{author}{\bibfnamefont{L.~M.} \bibnamefont{Johansen}},
  \bibinfo{journal}{Phys. Lett. A} \textbf{\bibinfo{volume}{366}},
  \bibinfo{pages}{374} (\bibinfo{year}{2007}{\natexlab{a}}).

\bibitem[{\citenamefont{Johansen}(2007{\natexlab{b}})}]{Johansen-Quantheosuccp%
roj:07}
\bibinfo{author}{\bibfnamefont{L.~M.} \bibnamefont{Johansen}},
  \bibinfo{journal}{Phys. Rev. A} \textbf{\bibinfo{volume}{76}},
  \bibinfo{pages}{012119} (\bibinfo{year}{2007}{\natexlab{b}}).

\bibitem[{\citenamefont{Wiseman}(2002)}]{Wiseman-Weakvaluquantraj:02}
\bibinfo{author}{\bibfnamefont{H.~M.} \bibnamefont{Wiseman}},
  \bibinfo{journal}{Phys. Rev. A} \textbf{\bibinfo{volume}{65}},
  \bibinfo{pages}{032111} (\bibinfo{year}{2002}).

\bibitem[{\citenamefont{Brunner et~al.}(2003)\citenamefont{Brunner, Acin,
  Collins, Gisin, and Scarani}}]{Brunner+AcinETAL-OptiTeleNetwWeak:03}
\bibinfo{author}{\bibfnamefont{N.}~\bibnamefont{Brunner}},
  \bibinfo{author}{\bibfnamefont{A.}~\bibnamefont{Acin}},
  \bibinfo{author}{\bibfnamefont{D.}~\bibnamefont{Collins}},
  \bibinfo{author}{\bibfnamefont{N.}~\bibnamefont{Gisin}}, \bibnamefont{and}
  \bibinfo{author}{\bibfnamefont{V.}~\bibnamefont{Scarani}},
  \bibinfo{journal}{Phys. Rev. Lett.} \textbf{\bibinfo{volume}{91}},
  \bibinfo{pages}{180402} (\bibinfo{year}{2003}).

\bibitem[{\citenamefont{Solli et~al.}(2004)\citenamefont{Solli, McCormick,
  Chiao, Popescu, and Hickmann}}]{Solli+McCormickETAL-FastLighSlowLigh:04}
\bibinfo{author}{\bibfnamefont{D.~R.} \bibnamefont{Solli}},
  \bibinfo{author}{\bibfnamefont{C.~F.} \bibnamefont{McCormick}},
  \bibinfo{author}{\bibfnamefont{R.~Y.} \bibnamefont{Chiao}},
  \bibinfo{author}{\bibfnamefont{S.}~\bibnamefont{Popescu}}, \bibnamefont{and}
  \bibinfo{author}{\bibfnamefont{J.~M.} \bibnamefont{Hickmann}},
  \bibinfo{journal}{Phys. Rev. Lett.} \textbf{\bibinfo{volume}{92}},
  \bibinfo{pages}{043601} (\bibinfo{year}{2004}).

\bibitem[{\citenamefont{Brunner et~al.}(2004)\citenamefont{Brunner, Scarani,
  Wegm\"uller, Legr\'e, and Gisin}}]{Brunner+ScaraniETAL-DireMeasSupeGrou:04}
\bibinfo{author}{\bibfnamefont{N.}~\bibnamefont{Brunner}},
  \bibinfo{author}{\bibfnamefont{V.}~\bibnamefont{Scarani}},
  \bibinfo{author}{\bibfnamefont{M.}~\bibnamefont{Wegm\"uller}},
  \bibinfo{author}{\bibfnamefont{M.}~\bibnamefont{Legr\'e}}, \bibnamefont{and}
  \bibinfo{author}{\bibfnamefont{N.}~\bibnamefont{Gisin}},
  \bibinfo{journal}{Phys. Rev. Lett.} \textbf{\bibinfo{volume}{93}},
  \bibinfo{pages}{203902} (\bibinfo{year}{2004}).

\bibitem[{\citenamefont{Williams and
  Jordan}(2008)}]{Williams+Jordan-WeakValuLeggIneq:08}
\bibinfo{author}{\bibfnamefont{N.~S.} \bibnamefont{Williams}} \bibnamefont{and}
  \bibinfo{author}{\bibfnamefont{A.~N.} \bibnamefont{Jordan}},
  \bibinfo{journal}{Phys. Rev. Lett.} \textbf{\bibinfo{volume}{100}},
  \bibinfo{pages}{026804} (\bibinfo{year}{2008}).

\bibitem[{\citenamefont{Ritchie et~al.}(1991)\citenamefont{Ritchie, Story, and
  Hulet}}]{Ritchie+StoryETAL-RealMeasWeakValu:91}
\bibinfo{author}{\bibfnamefont{N.~W.~M.} \bibnamefont{Ritchie}},
  \bibinfo{author}{\bibfnamefont{J.~G.} \bibnamefont{Story}}, \bibnamefont{and}
  \bibinfo{author}{\bibfnamefont{R.~G.} \bibnamefont{Hulet}},
  \bibinfo{journal}{Phys. Rev. Lett.} \textbf{\bibinfo{volume}{66}},
  \bibinfo{pages}{1107} (\bibinfo{year}{1991}).

\bibitem[{\citenamefont{Parks et~al.}(1998)\citenamefont{Parks, Cullin, and
  Stoudt}}]{Parks+CullinETAL-ObsemeasoptiAhar:98}
\bibinfo{author}{\bibfnamefont{A.~D.} \bibnamefont{Parks}},
  \bibinfo{author}{\bibfnamefont{D.~W.} \bibnamefont{Cullin}},
  \bibnamefont{and} \bibinfo{author}{\bibfnamefont{D.~C.}
  \bibnamefont{Stoudt}}, \bibinfo{journal}{Proc. R. Soc. Lond. A}
  \textbf{\bibinfo{volume}{454}}, \bibinfo{pages}{2997} (\bibinfo{year}{1998}).

\bibitem[{\citenamefont{Pryde et~al.}(2005)\citenamefont{Pryde, O�Brien, White,
  Ralph, and Wiseman}}]{Pryde+OBrienETAL-MeasQuanWeakValu:05}
\bibinfo{author}{\bibfnamefont{G.~J.} \bibnamefont{Pryde}},
  \bibinfo{author}{\bibfnamefont{J.~L.} \bibnamefont{O�Brien}},
  \bibinfo{author}{\bibfnamefont{A.~G.} \bibnamefont{White}},
  \bibinfo{author}{\bibfnamefont{T.~C.} \bibnamefont{Ralph}}, \bibnamefont{and}
  \bibinfo{author}{\bibfnamefont{H.~M.} \bibnamefont{Wiseman}},
  \bibinfo{journal}{Phys. Rev. Lett.} \textbf{\bibinfo{volume}{94}},
  \bibinfo{pages}{220405} (\bibinfo{year}{2005}).

\bibitem[{\citenamefont{Wang et~al.}(2006)\citenamefont{Wang, Sun, Zhang,
  Jian-Li, Huang, and Guo}}]{Wang+SunETAL-Expedemomethreal:06}
\bibinfo{author}{\bibfnamefont{Q.}~\bibnamefont{Wang}},
  \bibinfo{author}{\bibfnamefont{F.-W.} \bibnamefont{Sun}},
  \bibinfo{author}{\bibfnamefont{Y.-S.} \bibnamefont{Zhang}},
  \bibinfo{author}{\bibnamefont{Jian-Li}},
  \bibinfo{author}{\bibfnamefont{Y.-F.} \bibnamefont{Huang}}, \bibnamefont{and}
  \bibinfo{author}{\bibfnamefont{G.-C.} \bibnamefont{Guo}},
  \bibinfo{journal}{Phys. Rev. A} \textbf{\bibinfo{volume}{73}},
  \bibinfo{pages}{023814} (\bibinfo{year}{2006}).

\bibitem[{\citenamefont{Brun et~al.}(2008)\citenamefont{Brun, Di\'{o}si, and
  Strunz}}]{Brun+DiosiETAL-Testweakmeastwo-:08}
\bibinfo{author}{\bibfnamefont{T.~A.} \bibnamefont{Brun}},
  \bibinfo{author}{\bibfnamefont{L.}~\bibnamefont{Di\'{o}si}},
  \bibnamefont{and} \bibinfo{author}{\bibfnamefont{W.~T.}
  \bibnamefont{Strunz}}, \bibinfo{journal}{Phys. Rev. A}
  \textbf{\bibinfo{volume}{77}}, \bibinfo{pages}{032101}
  (\bibinfo{year}{2008}).

\bibitem[{\citenamefont{Garretson et~al.}(2004)\citenamefont{Garretson,
  Wiseman, Pope, and Pegg}}]{Garretson+WisemanETAL-uncerelawhicexpe:04}
\bibinfo{author}{\bibfnamefont{J.~L.} \bibnamefont{Garretson}},
  \bibinfo{author}{\bibfnamefont{H.}~\bibnamefont{Wiseman}},
  \bibinfo{author}{\bibfnamefont{D.~T.} \bibnamefont{Pope}}, \bibnamefont{and}
  \bibinfo{author}{\bibfnamefont{D.~T.} \bibnamefont{Pegg}},
  \bibinfo{journal}{J. Opt. B: Quantum Semiclass. Opt.}
  \textbf{\bibinfo{volume}{6}}, \bibinfo{pages}{S506} (\bibinfo{year}{2004}).

\bibitem[{\citenamefont{Johansen and
  Mello}(2008)}]{Johansen+Mello-QuanMechSuccMeas:08}
\bibinfo{author}{\bibfnamefont{L.~M.} \bibnamefont{Johansen}} \bibnamefont{and}
  \bibinfo{author}{\bibfnamefont{P.~A.} \bibnamefont{Mello}},
  \bibinfo{journal}{Phys. Lett. A} \textbf{\bibinfo{volume}{372}},
  \bibinfo{pages}{5760} (\bibinfo{year}{2008}).

\bibitem[{\citenamefont{{von Neumann}}(1955)}]{Neumann-MathFounQuanMech:55}
\bibinfo{author}{\bibfnamefont{J.}~\bibnamefont{{von Neumann}}},
  \emph{\bibinfo{title}{Mathematical Foundations of Quantum Mechanics}}
  (\bibinfo{publisher}{Princeton University Press},
  \bibinfo{address}{Princeton}, \bibinfo{year}{1955}).

\bibitem[{\citenamefont{Johansen}(2004)}]{Johansen-WeakMeaswithArbi:04}
\bibinfo{author}{\bibfnamefont{L.~M.} \bibnamefont{Johansen}},
  \bibinfo{journal}{Phys. Rev. Lett.} \textbf{\bibinfo{volume}{93}},
  \bibinfo{pages}{120402} (\bibinfo{year}{2004}).

\end{thebibliography}
\end{document}